%% file: mass_super_v3.tex
\documentclass[reprint, aps, prl, superscriptaddress,twocolumn,amsmath,amssymb,nofootinbib]{revtex4-1}

\usepackage{graphicx}
\usepackage{graphics}
\usepackage{color}
\usepackage{amsmath,amssymb,amsthm,mathtools}
\usepackage{epstopdf,cancel,ulem}
\usepackage{epsf,latexsym,bbm,euscript}
\usepackage[colorlinks]{hyperref}

\input{thesis_defs}

\usepackage{comment} 
\usepackage{color} 
\usepackage{cancel} 

\begin{document}


\title{Puzzling out the mass-superselection rule}

\author{Magdalena Zych}
\affiliation{Centre for Engineered Quantum Systems, School of Mathematics and Physics, The University of Queensland, St Lucia, Queensland 4072, Australia}
 \author{Daniel M. Greenberger}
\affiliation{Department of Physics, City College of the City University of New York, 10031, New York, USA}

\begin{abstract}
Mass-superselection rule (MSR) states that in the non-relativistic quantum theory superpositions of states with different masses are unphysical. While MSR features even in textbooks, its validity, physical content and consequences remain debated. Its original derivation is known to be inconsistent, while a consistent approach  does not yield a superselection rule. Yet, superpositions of masses do not seem to be present in Newtonian physics. Here we offer a resolution of the MSR puzzle.  Crucial for the result is the understanding of two issues:  how the notion of a mass parameter arises from the fundamentally dynamical relativistic notion of mass-energy, and what is the correct Newtonian limit of relativistic dynamics of composite particles. The result explains the physical content of the MSR without invoking any non-standard physics and clarifies the relation between MSR and a formalism describing relativistic composite particles, developed for studying relativistic effects in table top experiments. 
\end{abstract}
\maketitle

The status and physical meaning of the mass superselection rule (MSR) have been discussed from a variety of perspectives. It has been considered that some dynamical process could be suppressing superpositions of states associated with different masses~\cite{giulini2000states, ref:GiuliniBook, Annigoni2013}, it has been proposed  that in the non-relativistic quantum theory spacetime is not Newtonian~\cite{Hernandez-Coronado2012}, while in the context of quantum reference frames it has been questioned whether mass-superpositions are at all physical~\cite{Pereira:2015Gal}.   
Finally,  proposals were put forward to simulate dynamics of mass-superpositions with quantum optical setups in order to test MSR experimentally~\cite{Longhi:18}. {On the other hand a fully consistent formalism has been developed where  mass and proper time are conjugate dynamical variables~\cite{Greenberger:1970_1, Greenberger:1970_2, Greenberger:1974, Greenberger:2001, HERNANDEZCORONADO20132293}. This challenges the above  approaches and shows that neither physical meaning nor validity of MSR  are fully understood.}
 
Here we point out  a natural explanation of the physical content of the MSR, answering the question: Why we do not observe mass-superpositions in the non-relativistic physics? The key to the resolution is the correct definition of a mass parameter for a composite system and of the non-relativistic limit of its dynamics. The result also offers an interpretation of the Newtonian theory of particles with dynamical masses in terms of low-energy relativistic particles with dynamical internal degrees of freedom. 
 Our results  are relevant to the growing body of works studying relativistic effects in low-energy quantum systems~\cite{Zych:2011, pikovskiuniversal2015, bushev2016single,  Pang:2016, Orlando2017, sonnleitner2017will,  zych2015quantumEEP, paige2018quantum}, and clarify why MSR does not invalidate the analyses, contrary to some arguments~\cite{bonder2016questioning}.  

 \paragraph{Mass-superselection rule.--}
Bargman's original argument for the mass-superselection rule~\cite{Bargmann:1954}, see also~\cite{Greenberger:1979_EGT, Giulini:1996, Greenberger:2001}, 
arises for a quantum theory with Galilean symmetry. Galilei group $\mathcal G$ comprises spatial translations $\vec a$; temporal translations $b$; boosts  $\vec w$; and rotations $R$. 
A group element $g_{R, \vec w, \vec a, b}\in\mathcal G$ acts on coordinates as $ g(\vec x, t) = (R\vec x+\vec w t +\vec a, t+b)$. In Hilbert space, the action of $g(\vec x, t)$ is represented by unitary operators $U_{g_{\alpha}}=e^{-\frac{i}{\hbar} G\alpha}$ with $G$ being a generator of the transformation. In particular, the generator of translations is the momentum $\vec p$ and of  boosts is $\vec K=m \vec x-\hbar t \vec p$, where $\vec x$ is the position operator and $m$ is the mass parameter. The crux of the argument is the observation that boosts and translations in the Galilei group commute, while their unitary representations do not 
and a unitary representation of the Galilei group is projective \cite{WeinbergQFT:1995, Bargmann:1954, Levy-Leblond:1963, Giulini:1996}. As a result, the following sequence of transformations composed of translations $g_{\pm\vec a}$ and boosts $g_{\pm\vec w}$ yields an identity $id_{\mathcal G}$ in $\mathcal{G}$: 
\be
{g}_{-\vec a}{g}_{-\vec w}{g}_{\vec a}{g}_{\vec w}=id_{\mathcal G},
\ee{Bargmann_loop} 
while its unitary representation is not a identity in the Hilbert space of the particle
\be
U(g_{-\vec a})U({g}_{-\vec w})U({g}_{\vec a})U({g}_{\vec w})=e^{-im\vec a \vec w} I_{int}.
\ee{loop_unitaries}
Applying the transformation \eqref{loop_unitaries} to a superposition state of two masses $m$ and $m'$ predicts a relative phase $e^{i\vec w\vec a(m-m')}$ between the states, and thus a {different state} to the original one, unless $m = m'$. However, since the operation  \eqref{Bargmann_loop}  represents identity in $\mathcal G$ it  {cannot alter physical states}. Hence the superselection rule stating that superpositions of states with different masses do not represent physical states in a Galilei-invariant theory. 

\paragraph{Inconsistency of Bargman's result.--}
Bargman's argument has been criticised as inconsistent on the following grounds: \textit{(i)} In a theory with mass being a parameter there are no states for a single particle that are associated with different masses~\cite{Giulini:1996}; \textit{(ii)} 
 Newtonian theory is a limit of the relativistic theory where no superselection rule arises~\cite{WeinbergQFT:1995}, 
but how could a greater (Lorentz) symmetry give rise to a selection rule only in its limit~\cite{greenberger2010conceptual}? (A greater symmetry should lead to more restrictions on the state space.)
Furthermore, the phase in eq.~\eqref{loop_unitaries} has a direct physical interpretation in terms of time dilation~\cite{Greenberger:1979_EGT, Greenberger:2001}, 
which becomes evident when looking at an extended Galilei transformation~\cite{Greenberger:1979_EGT}: $\vec x'=\vec x-\vec\xi(t)$, $t=t'$ with arbitrary $\vec\xi(t)$. The  non-relativistic Schr\"{o}dinger equation $(\vec{p\,}^2/2m)\psi(\vec x, t)=i\hbar \dot\psi(\vec x, t)$, where $\dot\psi:=d\psi/d t$, in the above primed coordinates becomes 
$i\hbar\dot\varphi=\left( \frac{\vec p^{\,\prime2}}{2m}+m\ddot{\vec\xi}\cdot\vec x'\right)\varphi$, 
where $\psi(\vec x, t)=e^{if(\vec x',t)}\varphi(\vec x',t)$ and $f(\vec x',t)=m(\dot{\vec\xi}\cdot\vec x'+\int \!dt\,\dot\xi^2/2)/\hbar$. 
Generalising the sequence of boosts and shifts to an arbitrary closed path, $\vec\xi(0)=\vec\xi(T)=0$, the state in the primed frame at the end acquires a phase $\psi'(T)=\exp\{{(im/\hbar)\int_0^T\!dt\,\dot\xi^2/2}\}\psi(T)$. Proper time elapsing during this round-trip as measured in the primed coordinates is $T'=\int_0^T\!dt({1-\dot\xi^2/c^2})^{1/2}$ and the above phase factor is { 
given by time dilation $\Delta\tau =T-T'$ between the two frames $mc^2\Delta\tau= m\int_0^T\!dt\,{\dot\xi^2/2} +\mathcal{O}(c^{-4})$.} These observations will be crucial for our resolution. 

\paragraph{Newtonian particles with a dynamical mass.--}
To investigate the problem in a consistent manner, Newtonian theory has been extended to include mass as a dynamical operator $M$. The simplest extension yields a Hamiltonian~\cite{Greenberger:1970_1, Greenberger:1970_2, Greenberger:1974, Giulini:1996, Hernandez-Coronado2012, HERNANDEZCORONADO20132293} 
\be
H_{dm}=\frac{\vec p^{\,2}}{2M}+M\Phi(x),
\ee{dyn_mass_Newt}
where we have included the non-inertial term $\ddot{\vec\xi}\cdot\vec x'$ as a gravitational potential $\Phi(x)$, in light of the equivalence principle. Since the Newtonian limit of relativistic dynamics  contains the term $mc^2$, the corresponding term $Mc^2$ is often included in \eqref{dyn_mass_Newt}. The argument we outline below works in either case. 

The surprising finding is that no MSR arises in this dynamical-mass extension of the Newtonian dynamics, see ref.~\cite{Giulini:1996} for details.  
The reason is that the symmetry of eq.~\eqref{dyn_mass_Newt} is not $\mathcal{G}$ but its central extension  $\mathcal{\tilde{G}}$, with $M$ as the central element\footnote{Central extension of a group is a group whose quotient by the one-parameter subgroup, here generated by $M$,  is isomorphic to the original group, and where this subgroup commutes with all other group generators.}. The new group elements $\tilde g_{\alpha, R, \vec w, \vec a, b}\in \mathcal{\tilde{G}}$  implement the action of the Galilei group on the spacetime coordinates, as well as shifts along an additional coordinate $q$ associated with the mass. One can thus write $\tilde g\equiv(\alpha, g)$ whose action on all the coordinates reads \cite{Giulini:1996} $\tilde g(q, \vec x, t) = (q + \alpha - \vec w  R\vec x - \frac{1}{2}\vec w^2t,  R\vec x+\vec w t +\vec a, t+b)$. Crucially, the chain of transformations  \eqref{Bargmann_loop}
in $\mathcal{\tilde{G}}$ does not result in an identity but in a shift of the internal coordinate by $\vec w\vec a$
\be
\tilde{g}_{-\vec a}\tilde{g}_{-\vec w}\tilde{g}_{\vec a}\tilde{g}_{\vec w}
=(\vec w\vec a, id_{\mathcal G}).
\ee{central_G_loop}
The unitary operators implementing $\mathcal{\tilde{G}}$ are  the same as in the  Galilei group, with the replacement of $m\rightarrow M$, where $M$ is a generator of translations along $q$. Therefore, the unitary representation of the loop remains as in eq.~\eqref{loop_unitaries} with $m$ replaced by $M$. 
Instead of the pure phase $e^{-im\vec a \vec w}$ we thus have an operator $e^{-iM\vec a \vec w}$ implementing the same shift along $q$ as found  in eq.~\eqref{central_G_loop}.

We can now appreciate the essence of the MSR impasse: mass superpositions are not constrained by any superselection rule in a theory which actually includes different mass states that could be superposed. But if indeed there are no constraints on mass-superpositions, why aren't they ubiquitous in Newtonian physics? On the other hand, superposing masses unavoidably yields proper time effects, which have no interpretation in a non-relativistic theory.
Below we present our resolution to the problem, structured along the lines suggested in~\cite{Giulini:1996}:  find the system for which the states with different masses can actually be prepared; and address the question what is then measurable.

\paragraph{Relativistic composite particles.--}
The extended Galilei transformation hints that the resolution of the MSR puzzle must include the notion of proper time. The failure of the dynamical-mass extension of the Newtonian framework to yield MSR hints that dynamical mass should be taken seriously. 
%
We therefore begin with a fully relativistic theory of particles with internal degrees of freedom (DOFs)~\cite{Zych:2011, pikovskiuniversal2015, zych2015PhD, zych2018gravitational}. 

The square of the relativistic four momentum $p^\mu$, $\mu=0,..,3$ is an invariant quantity describing particle's energy in the rest frame \cite{WeinbergGR}  $\r c^2=-\sum p^{\mu}g_{\mu\nu}p^{\nu}$, where $g_{\mu\nu}$ is the metric tensor with signature $(-, +,+,+)$, and  $c$ is the speed of light. For a composite system, e.g.~an atom,  $\r$ comprises not only the sum of the masses of the constituents but also their binding and kinetic energies -- it is a fundamentally dynamical quantity describing internal DOFs. 
In an arbitrary reference frame, the total energy is $H\equiv cp_0$ and for a static symmetric metric reads 
\be
H= \sqrt{-\go(c^2\pj{}\Pj{} +\r^2)},
\ee{rel_comp_H}
where $p_jp^j\equiv\sum_{i,j=1,2,3}p^ig_{ij}p^j$; For a field-theory derivation see~\cite{zych2015PhD, Castro:2017clocks,  zych2015quantumEEP,Anastopoulos_2018}, for derivation as a limit of an N-particle bound system see~\cite{zych2018gravitational}. Locally, the symmetry of eq.~\eqref{rel_comp_H}  is the  central extension of the Poincar\'e group, with $\r$ being  the central element.  $\r$ is a generator of internal dynamics and  
commutes with all other generators. The  Poincar\'e algebra is otherwise unchanged compared to the case for a structureless particle where $m$ is a parameter labelling the representation. 
Such a central extension is `trivial', it is a product of the Poincar\'e and  of the internal symmetry group. 

We now seek the non-relativistic limit of eq.~\eqref{rel_comp_H}. The usual approach is to take a low-energy limit (small velocities and weak gravity) 
\be
H_{le}=\r+ \frac{{\vec p\,}^2c^2}{2\r} + \r\frac{\Phi(x)}{c^2},
\ee{low_en_H}
{where $\vec p=(p^1, p^2, p^3)$ is the three-momentum. Crucially the dynamics of the centre of mass (CM) given by  $H_{le}$
 and by $H_{dm}$  is fully equivalent -- as can  directly seen by replacing the internal energy in eq.~\eqref{low_en_H} with the dynamical mass in appropriate units: $\r\rightarrow Mc^2$.} Below we show that  $H_{le}$, eq~\eqref{low_en_H}, and therefore also $H_{dm}$, \eqref{dyn_mass_Newt} do not define a non-relativistic theory. 

Time evolution of an internal observable $a$ under $H_{le}$ is described by $\dot a=i[ H_{le}, a]/\hbar=:\omega$ where
\be
\omega=\omega_0\left(1-\frac{{\vec v\,}^2}{2c^2}+\frac{\Phi(x)}{c^2}\right),
\ee{time_dil_int}
 where $\vec v=\partial H_{le}/\partial {\vec p}$ 
 and $\omega_0:=i[\r, a]/\hbar$. Note that $\omega_0$ is the rest frame speed of internal dynamics and describes time evolution with respect to proper time $\tau$, i.e.~$\omega_0=\frac{da}{d\tau}$. To lowest post-Newtonian order $d\tau=(1-\frac{\vec v^2}{2c^2}+\frac{\Phi(x)}{c^2})dt$ and eq.~\eqref{low_en_H} thus includes the lowest order time dilation effects:  the velocity-dependent term describes the special relativistic time dilation and the potential-dependent term describes the gravitational time dilation, see also \cite{Zych2016, paige2018quantum}. 
 
The first essential observation is that for composite particles, one needs to better define when a theory is non-relativistic.
{We propose the following:} a non-relativistic limit should give rise to Euclidean notion of spacetime, with global time. This is the case in eq.~\eqref{time_dil_int} when 
$\omega\approx\omega_0$. {To understand under what conditions this happens it is instructive to split the rest energy $\r$ into a static part $E_{0}\!\cdot\!I_{int}$,  where $I_{int}$ is the identity operator on the internal DOFs (which we skip hereafter) and the remaining dynamical part $H_0:=\r-E_{0}$, so that $\r\equiv E_{0} +H_{0}$. We can now take the limit $H_0\ll E_{0}$ of eq.~\eqref{low_en_H}, which to lowest order in $1/c^2$ yields} 
\be
\begin{split}
H_{le}&\approx  E_{0}+\frac{{\vec p\,}^2c^2}{2E_{0}} + E_{0}\frac{\Phi(x)}{c^2}+H_0\\
+&H_0\bigg(\!-\!\frac{{\vec p\,}{^2}c^2}{2E_{0}^2}+\frac{\Phi(x)}{c^2}\bigg).
\end{split}\ee{lowe_split}
The first three terms do not contribute to internal dynamics and 
the term $H_0$ alone gives universal internal evolution, independent of the CM. The notion of a global time is therefore recovered when the remaining terms, the second line of eq.~\eqref{lowe_split}, 
are negligible. When these terms  are absent, the rate of internal dynamics is independent of the  velocity or gravitational potential difference between the rest frame of the particle and the frame with time coordinate $t$. 

The second essential observation is that we do not automatically have a notion of a mass-parameter for a relativistic composite system -- we only have a dynamical quantity $\r$, {the energy in the rest frame of the system~\cite{Greenberger:2001}}, which in the low-energy limit  defines all the mass-energies: the rest mass-energy, inertia and weight, see eq.~\eqref{low_en_H}.  Eq.~\eqref{lowe_split} offers a natural definition of the mass \textit{parameter}  as the static part of $\r$, i.e.~$m:=E_{0}/c^2$. Incorporating this into eq.~\eqref{lowe_split} and taking $H_0/mc^2\ll I$ (the first observation above) gives the correct non-relativistic Hamiltonian of a composite system 
\be
mc^2+H_{0}+ \frac{\vec{p}^{\,2}}{2m} + m\Phi(x).
\ee{fullNewt}
Note that the dynamical mass-energy is not entirely suppressed but survives in the rest energy term, resulting in the familiar expression for a total energy of a non-relativistic composite system -- where internal energy $H_{0}$ simply adds to the CM kinetic and potential energies. 

\paragraph{Composite particles and MSR.--}
We can now analyse what happens in the correct Newtonian limit \eqref{fullNewt} for a superposition state of two mass-energies $\ket{M_1}+\ket{M_2}$, where $\ket{M_i}$, $i=1,2$ is the eigenstate of $M=\r/c^2\equiv mI_{int}+H_0/c^2$ with the eigenvalue $M_i$. We have $M(\ket{M_1}+\ket{M_2}) = m(\ket{M_1} +\ket{M_2}) +E_1/c^2\ket{M_1}+E_{2}/c^2\ket{M_2}$, where $E_i$ is the internal energy $H_{0}\ket{M_i} = E_i\ket{M_i}$.  
The state is a superposition of mass-energies, but it has a well-defined mass, since by definition the Newtonian mass is an operator proportional to identity, $mI_{int}$. 
Mass-energy superpositions are therefore neither forbidden nor lead to superpositions of Newtonian masses: In the correct Newtonian limit the dynamical part of the mass-energy $H_0$ is negligible in the inertial and gravitational potential energy terms, where only the static part $mI_{int}$ contributes. It is this static part which we recognise as ``the mass'' in the non-relativistic physics. 

Allthough in the Newtonian limit the state $\ket{M_1}+\ket{M_2}$ has a fixed value $m$ of inertia and weight, it is in a superposition of rest mass-energies, as the dynamical part  of the mass-energy survives as the additive internal energy $H_0$ in eq~\eqref{fullNewt}. 

\paragraph{How does $\mathcal{G}$ emerge from $\mathcal{\tilde{G}}$?}
As anticipated~\cite{Giulini:1996}, understanding MSR also explains how $\mathcal{G}$ emerges from $\mathcal{\tilde{G}}$ as a symmetry in non-relativistic physics. 
Note first that non-relativistic limit of the Poincar\'{e} group (its In\"{o}n\"{u}-Wigner contraction \cite{Inoenue:1952, WeinbergQFT:1995}) for structureless particles is the central extension of the Galilei group and not the Galilei group \cite{WeinbergQFT:1995}. Consequently,  the low-energy limit of the central extension of the Poincar\'{e} group, with $mc^2\rightarrow\r$, is the central extension of the Galilei group with $\r/c^2$ as the centre \cite{zych2015PhD}. This can be seen as a physical reason why $\mathcal{\tilde{G}}$ arises as a symmetry of ``Newtonian'' particles with dynamical masses, found in ref.~\cite{Giulini:1996}.
%
%
%

We have just shown that a consistent non-relativistic limit for composite particles is obtained when the inertial and gravitational mass-energies are effectively a parameter. In the unitary representation of the boost we are then left with a parameter $m$ instead of the operator $M$, since it is the inertial mass which is relevant here.  The commutator between the boost and the translation generators thus  becomes $pK-Kp\rightarrow m$. The resulting symmetry group is a  product of the central extension of the Galilei group with a parameter $m$ at the centre and a one parameter group of internal symmetries generated by $H_0$. 
%
%
However, transformations resulting from such a central extension of the Galilei group (with a parameter $m$ in the centre) are indistinguishable from those originating from the Galilei group itself. The two symmetries are empirically indistinguishable~\cite{WeinbergQFT:1995, Bargmann:1954, Levy-Leblond:1963, Giulini:1996}, 
both in the quantum and in the classical case
\footnote{Te extended Galilei group already appears as a symmetry in classical phase space:  commutators of group elements in the quantum case and their Poisson brackets in the classical case are the same up to the imaginary unit. Consequently, central extension of the Galilei group also appears as a symmetry in a classical theory where the mass is taken to be dynamical \cite{Giulini:1996}.}

\paragraph{What is measurable in the Newtonian framework?}
For a non-relativistic limit of any theory to be meaningful, we have to assume that measurements can only have a finite precision. If we could measure internal states arbitrarily precisely,  time dilation of their evolution would never be negligible.  Experiments with atomic clocks have already measured time dilation between clocks with relative speed of $\!\sim\!10$m/s, and at a height difference of  $\!\sim\!30$cm~\cite{Wineland:2010}. State of the art clocks could even  measure time dilation due to relative velocities  $\!\sim\!30$cm/s and  height difference $\!\sim\!2$cm~\cite{Ye2015}. This further illustrates that for systems with internal DOFs one cannot meaningfully define the Newtonian limit only in terms of the CM. 

The lack of signatures of dynamical masses in Newtonian physics is  therefore not due to the lack of observables that can measure mass-energy in superposition, but due to the fact that the Newtonian limit is defined as the limit where dynamical mass-energy contributions to inertia and weight are negligible. Consider Rabi oscillations between internal states of an atom~\cite{foot2005atomic}. They demonstrate coherence between different eigenstates of $H_0$, and thus of $M$. But unless their frequency is very high, the physical effects arising from the dynamical part of the atom's inertia and weight are negligible. On the other hand, the above mentioned time dilation measured with atomic clocks directly verifies the dynamical nature of inertia and weight as described by $H_{le}$ (these efffects are fully explained by eq.~\eqref{time_dil_int}). We usually do not think of time dilation effects as demonstrating mass-superpositions -- nor as effects violating MSR -- although $H_{le}$ with $\r\rightarrow Mc^2$ is operationally the same as $H_{dm}$, with the added rest mass term, as in refs~\cite{Greenberger:1970_1, Greenberger:1970_2, Greenberger:1974, Greenberger:2001, HERNANDEZCORONADO20132293}.



\paragraph{Einstein Equivalence Principle.--}
 The notion of a mass parameter emerges in the non-relativistic limit {by splitting  the mass-energy into two separate quantities: mass and internal energy}.\footnote{Even more broadly,  the masses of most standard model particles arise as a static energy term -- where the energy is that of the Higgs field.}
 The Newtonian limit thus  breaks two of the three constituents of the Einstein Equivalence Principle~\cite{will:2014xja}: Local Lorentz Invariance and Local Position Invariance, where the former requires equality of rest mass-energy and inertia and the latter -- of rest mass-energy and weight.
Indeed, in the non-relativistic limit \eqref{fullNewt} rest mass-energy is essentially unchanged ($H_{0}$ and $\r$ only differ by an unobservable constant $E_0$), whereas inertia and weigh are given by $E_{0}/c^2$. This preserves the validity of the remaining part of the EEP, the Weak Equivalence Principle. For discussion and experimental tests of the EEP for composite quantum particles see refs~\cite{Orlando:2016EEP,  rosi2017quantum, geiger2018proposal, zych2015quantumEEP, Anastopoulos_2018}. 

\paragraph{Conclusion.--}
Lack of mass-superpositions in Newtonian physics is a consequence of the operational definition of the mass-parameter and of the non-relativistic limit of dynamics of composite particles. It does not require any restriction on kinematics. 
{The problem with MSR has been rooted in a presumption that promoting mass to a dynamical variable in a Newtonian Hamiltonian still yields a non-relativistic theory~\cite{Giulini:1996, giulini2000states, ref:GiuliniBook, Annigoni2013, HERNANDEZCORONADO20132293, Hernandez-Coronado2012,Pereira:2015Gal}. 
However, promoting mass to a dynamical quantity is  operationally equivalent to incorporating relativistic mass-energy equivalence, which brings in the time dilation effects~\cite{Greenberger:1970_1,Greenberger:1970_2,  Greenberger:2001, Zych:2011, zych2015quantumEEP}.} In fact, mass-energy equivalence has been used by Einstein to derive the gravitational time dilation~\cite{Einstein:1911}.
{The symmetry of the dynamical-mass framework comprises Galilean transformations for spacetime coordinates and translations along a new coordinate associated with the dynamical mass~\cite{Giulini:1996,HERNANDEZCORONADO20132293, Hernandez-Coronado2012}. The low-energy relativistic theory is instead invariant under central extension of the Galilei group with mass-energy as the central element, which is the Lorentz symmetry up to $1/c^2$~\cite{zych2015PhD}, and proper time takes the role of the additional coordinate~\cite{Greenberger:1970_1,Greenberger:1970_2,  Greenberger:2001}. }

Independently of its interpretation, the regime of low-energy particles with dynamical mass-energy has its own symmetry and phenomenology, and can be studied fully in its own right. It has already allowed exploring proper time effects in unstable~\cite{Greenberger:1974} and interfering quantum particles~\cite{Zych:2011, margalit2015self, bushev2016single, Pang:2016, Orlando2017},  it was used to assess the limits to the notion of an ideal clock~\cite{paige2018quantum, Sinha_2014} and to the notion of time~\cite{Castro:2017clocks},  and to study the role of mass-energy equivalence in atom-light interactions~\cite{sonnleitner2017will, sonnleitner2018mass}. The approach has further enabled a quantum formulation of the EEP for composite particles~\cite{zych2015quantumEEP} and can shed light on the role of proper time in quantum-to-classical transition~\cite{pikovskiuniversal2015,ref:AB,Korbicz2016,PikovskiTime2017}.



\paragraph{Acknowledgment.--} 
M.Z acknowledges  Australian Research Council (ARC) DECRA grant DE180101443 and ARC Centre EQuS CE170100009. This publication was made possible through the support of a grant from the John Templeton Foundation. The opinions expressed in this publication are those of the authors and do not necessarily reflect the views of the John Templeton Foundation. The authors acknowledge the traditional owners of the land on which the University of Queensland is situated, the Turrbal and Jagera people. 

\bibliographystyle{linksen}
\bibliography{bibliomass}
\end{document}

%% file: thesis_defs.tex
\newcommand{\pj}[1]{p_{#1j}}
\newcommand{\Pj}[1]{p_{#1}^j}

\newcommand{\Xip}[1]{x_{#1}^{\prime i}}

\def\gij{g_{ij}}

\def\go{g_{00}}

\def\r{H_{r}}

\newcommand {\ket}[1]{\lvert \, #1\rangle}

\def\td{\dot\tau}

\newcommand{\be}{\begin{equation}}
\newcommand{\ee}[1]{\label{#1} \end{equation}}
\newcommand{\bbe}{\begin{equation*}}
\newcommand{\eee}{\end{equation*}}

\newcommand{\bs}{\begin{split}}
\newcommand{\es}{\end{split}}

\newcommand{\e}[1]{e^{{ \textstyle #1 }}}

\def\f12{\frac{1}{2}}

